# Variability of research performance across disciplines within universities in non-competitive higher education systems[1]

*Giovanni Abramo[a], Ciriaco Andrea D'Angelo[a,b], Flavia Di Costa[a,c]*

[a] Laboratory for Studies of Research and Technology Transfer
Institute for System Analysis and Computer Science (IASI-CNR)
National Research Council of Italy

[b] Department of Engineering and Management
University of Rome "Tor Vergata"

[c] Research Value S.r.l.

**Abstract**

Many nations are adopting higher education strategies that emphasize the development of elite universities able to compete at the international level in the attraction of skills and resources. Elite universities pursue excellence in all their disciplines and fields of action. The impression is that this does not occur in "non-competitive" education systems, and that instead, within single universities excellent disciplines will coexist with mediocre ones. To test this, the authors measure research productivity in the hard sciences for all Italian universities over the period 2004-2008 at the levels of the institution, their individual disciplines and fields within them. The results show that the distribution of excellent disciplines is not concentrated in a few universities: top universities show disciplines and fields that are often mediocre, while generally mediocre universities will often include top disciplines.

# 1. Introduction

Education, with the resulting new knowledge creation and transfer, is the lifeblood of socio-economic growth. Universities are thus a cornerstone of the knowledge-based society. The substantial proof of the essential role of research in innovation and growth (e.g. Adams, 1990; Rosenberg and Nelson, 1994; Mansfield, 1995; Griliches, 1998; Henderson et al., 1998) has clearly led many nations to invest in national scientific infrastructure and the higher education system. Such policies often pursue the development and strengthening of world-class universities (Mohrman et al., 2008; Deem et al., 2008), able to compete globally to attract talented students, faculty, and abundant funding. Salmi (2009) identified three possible strategies available to the decision maker who wishes to promote the birth of world-class universities:

  i) Strengthen a small number of existing universities to bring them to excellent level (*picking winners*), a strategy recently adopted by the German federal government, which has organized a national competition to award substantial extra funding.
  ii) Stimulate fusion of a certain number of institutions (*hybrid formula*) - this strategy is seen in France and Denmark, where individual universities and *grandes écoles* are exploring the feasibility of regional mergers.
  iii) Create world-class universities from scratch (*clean-slate approach*) – examples of this strategy are the series of Indian Institutes of Technologies, which implement a research policy agenda placing top priority on science and technology.

Abramo et al. (2012) have proposed a fourth strategy, suggested particularly for undifferentiated non-competitive higher education systems: the spin-off of elite universities staffed with top scientists from existing ones. While it is obvious there is no universal recipe or magic formula for creating a world-class university, it is certainly possible to identify some of their characteristic features, such as high concentrations of talent, ample availability of resources and flexible models of governance. Yet in spite of attempts to define a vision and shared developmental strategy at the European Union level (EU Commission, 2003), the policies pursued by the member nations differ greatly from one another. Unlike certain other countries (with the United States and other "Anglo-Saxon" nations leading), the current situation in Europe is often one of excess public control, inefficient governance and chronic insufficiency of funds (Veugelers and Van der Ploeg, 2008). In the area of university governance, the OECD (2007) reveals that the most important driver of modernization in a university system is competition. The growth of world-class universities is favored by regulatory environments that introduce and permit competitive mechanisms for the stimulation of continuous improvement and the pursuit of competitive advantage. Aghion et al. (2009) show that universities' performance is correlated with their autonomy and competitive environment. The level of competitiveness depends on cultural and other contextual factors, particularly the level of university autonomy, the type of financing and a government regulatory framework that is supportive in character. In competitive higher education systems, such as those observed in English speaking nations, the pursuit of competitive advantage has led to development of world-class universities that can attract, develop and retain highly-talented national and foreign faculty and students. The same institutions also obtain abundant public financing, private financing and donations, and attract venture capital and establishments of national and international high-tech companies in their territory, with resulting social and economic benefits. The competition factor has generated universities that are distinct in their quality of education and research, and thus in prestige, and that offer degrees and produce research results that also stand out for their social and market value. Mechanisms for stimulating competition can be varied in response to the context, but such competition will inevitably lead to more efficient selection and continuous improvement. These competitive mechanisms have developed quite naturally in the Anglo-Saxon nations, with long-standing evolution of national policies that favored the birth and development of a true higher-education market, while the excess of public control seen in many European nations has inhibited the initiation of true competitive mechanisms and led to the development of a generally undifferentiated higher education system that is unable to compete at a global level for access to economic and human resources (public and private funds; talented students, excellent faculty) (Auranen and Nieminen, 2010).

If the objective of government is to stimulate competition, and so lead to development of top universities that can compete internationally and produce the relevant socio-economic benefits, then some form of observation and monitoring of the situation is obviously useful. Other than monitoring performance itself, another indicator is the degree of concentration of performance within universities: the less that research performance is dispersed within universities then the greater is the probability of having top universities that are able to compete at the international level. In the case of Italy's non-competitive higher education system, Abramo et al. (2012) found that the distribution of research impact among academics is generally highly concentrated on certain individuals, and further that there is more variability within the individual institutions than there is between universities as a whole. This seriously hampers the possibility of having top universities that are able to compete at the international level, since no single university reaches the critical mass of top performers able to attract relevant human and financial resources. To follow up on these findings in this work we wish to verify if the universities that place highest in national rankings also succeed in placing at the top of the productivity rankings for the single disciplines in which they operate. This is in fact what occurs in elite universities from countries with competitive higher education systems. In spite of the methodological weaknesses[2] of certain international rankings (e.g. Times Higher Education Supplement[3], Shanghai Jiao Tong University Ranking[4] and QS World University Rankings[5]), these do provide a glimpse of such consistency in ranking. Taking the example of MIT, the institution as a whole inevitably places among the top in the global international rankings, but also for its individual subject areas of activity (Engineering & technology; Life sciences & medicine; Natural sciences; Social sciences & Management, as named in the QS rankings). The same phenomenon emerges at the next lower level: taking the example of Engineering & technology (again for the QS rankings), MIT still appears at top position in the world rankings for its individual constituent fields (Computer science and information systems, Chemical engineering, Civil engineering, Electrical engineering, Mechanical engineering). Our question is whether this situation is repeated

---

[2]The various weaknesses of these international rankings have been thoroughly examined and reported: see in particular Van Raan (2005).
[3]http://www.thes.co.uk/worldrankings/, last accessed on July 1, 2013
[4] http://www.arwu.org/, last accessed on July 1, 2013
[5] http://www.topuniversities.com/university-rankings/world-university-rankings, last accessed on July 1, 2013

in non-competitive higher education systems, meaning that the research performance of a university in its individual disciplines is more or less homogenous, or is in contrast highly variable. To respond to this research question we refer to the Italian university system, which has characteristics making it particularly suited to such an analysis. Most essentially it is absolutely non-competitive, but it also has another convenient and apparently unique characteristic: the fact that each academic is classified by research specialty in a unique field, named Scientific Disciplinary Sector, or SDS, of which there are 370 grouped in 14 University Disciplinary Areas, or UDAs. This classification scheme makes it possible to list universities research productivity ranking lists by fields (SDSs) and disciplines (UDAs), thus permitting an answer to the research question.

To better explain the context for the analysis, the next section further describes the key characteristics of the Italian university system. Section 3 illustrates the methodology adopted to evaluate research productivity at the various levels of analysis, while Section 4 reports the results. The paper closes with the authors' thoughts on the results and policy suggestions.

## 2. The Italian higher education system

In Italy, the Ministry of Education, Universities and Research (MIUR) recognizes a total of 96 universities as having the authority to issue legally-recognized degrees. Sixty-seven are public and generally multi-disciplinary universities, scattered throughout the nation. Six of them are *Scuole Superiori* (Schools for Advanced Studies), specifically devoted to highly talented students, with very small faculties and strictly limited enrollment numbers per degree program. The 29 private universities are small sized; 94.9% of faculty is employed in public universities (0.5% in Scuole Superiori). Public universities are largely financed by government through non-competitive allocation. Up to 2009, the core government funding (56% of universities' total income) was input oriented, i.e. distributed to universities in a manner intended to equally satisfy the needs of each and all, in function of their size and activities. It was only following the first national research evaluation exercise (VTR), conducted between 2004 and 2006, that a minimal share, equivalent to 3.9% of total income, was attributed by the MIUR in function of assessment of research and teaching quality.

Despite some intervention intended to grant increased autonomy and responsibilities to universities (Law 168 of 1989), the Italian higher education system is a long-standing and classic example of a public and highly centralized governance structure, with low levels of autonomy at the university level and a very strong role played by the central state.

In keeping with the Humboldtian university model, with its emphasis on the unity of teaching and research, there are no "teaching-only" universities in Italy, as all professors are required to carry out both research and teaching. Furthermore, the time to devote to teaching is established by law. All new personnel enter the university system through public examinations and career advancement can only proceed by further public examinations. Salaries are regulated at the centralized level and are calculated according to role (administrative, technical, or professorial), rank within role (for example: assistant, associate or full professor) and seniority. None of a professor's salary depends on merit. Moreover, as in all Italian public administration, dismissal of an employee for lack of productivity is unheard of.

The whole of these conditions create an environment and a culture that are completely non-competitive, yet flourishing with favoritism and other opportunistic behaviors that are dysfunctional to the social and economic roles of the higher education system. The overall result is a system of universities that are almost completely undifferentiated for quality and prestige, with the exception of the tiny Scuole Superiori and a very small number of the private special-focus universities. In the period 2004-2008, about 25% of professors had no impact on scientific progress, while 23% alone produced 77% of the overall scientific advancement (Abramo et al., 2011a). The problem is that this 23% of faculty is not concentrated in a limited number of universities, but is instead dispersed more or less uniformly among all Italian universities, along with the unproductive ones, so that no single institution reaches the critical mass of excellence necessary to develop as an elite university and compete at the international level (Abramo et al., 2012).

## 3. Methodology

### 3.1 Measuring research productivity

Research activity is a production process aimed at producing and diffusing new knowledge. The new-knowledge production function has a multi-input and multi-output character, which makes the productivity of the total production factors not easily measurable. There are two traditional approaches used by scholars to measure the total factor productivity: parametric and non-parametric techniques. Parametric methodologies are based on the a priori definition of the function that can most effectively represent the relationship between input and output of a particular production unit. The purpose of non-parametric methods, on the other hand, is to compare empirically measured performances of production units (commonly known as Decision Making Units, DMUs), in order to define an "efficient" production frontier, comprising the most productive DMUs. The reconstruction of that frontier is useful to assess the inefficiency of the other DMUs, based on minimum distance from the frontier[6].

The measure of total factor productivity requires information on the different production factors by unit of analysis. Instead of total factor productivity, most often research administrators are interested in measuring and comparing simply labor productivity, i.e. the value of output per unit value of labor, normalized by all other production factors. When measuring labor productivity, if there are differences in the production factors (capital, scientific instruments, materials, etc.) available to each scientist then one should normalize by them. Unfortunately, in Italy relevant data are not available at individual level. We assume then that resources available to researchers within the same field are the same. A further assumption is that the hours devoted to research are more or less the same for all researchers. As we saw in the previous section, these assumptions are fairly well satisfied in the Italian higher education system, which is mostly public and not competitive.

It has been shown (Moed, 2005) that in the so-called hard sciences, the prevalent form of codification for research output is publication in scientific journals. Such databases as Scopus and Web of Science (WoS) have been extensively used and tested in bibliometric analyses, and are sufficiently transparent in terms of their content and coverage. As a proxy of total output in the hard sciences, we can thus simply consider publications indexed in either WoS or Scopus[7]. With this proxy, those publications that are not censused will inevitably be ignored. This approximation is considered acceptable in the hard sciences, although not for the arts, humanities and a good part of the social science fields. As proxy of the value of output we adopt the number of citations for the researcher's publications. Because the intensity of publication varies across fields (Garfield, 1979; Moed et al., 1985; Butler, 2007), in order to avoid distortions in productivity rankings, we compare researchers within the same field (SDS). It is very possible though that researchers belonging to a particular scientific field will also publish outside that field. Because citation behavior varies by field, we standardize the citations for each publication with respect to the mean of the distribution of citations for all the Italian cited publications of the same year and the same WoS subject category. Because research projects frequently involve a team of researchers, which shows in co-authorship of publications, productivity measures then need to account for the fractional contributions of single authors to outputs. The contributions of the individual co-authors to the achievement of the publication are not necessarily equal, and in some fields the authors signal the different contributions through their order in the byline. The conventions on the ordering of authors for scientific papers differ across fields (Pontille, 2004; RIN, 2009), thus the fractional contribution of the individuals must be weighted accordingly. Following these lines of logic, in our view productivity indicators based on full counting or "straight" counting (where only the first author or the corresponding author receive full credit and all others receive none) are invalid. The same invalidity applies to indicators based on equal fractional counting in fields where co-author order in the byline has recognized meaning. Finally, accounting for the cost of labor would require knowledge of the cost of each researcher, information that is usually unavailable for reasons of privacy. In the Italian case we have resorted to a proxy. In the Italian university system, salaries are established at the national level and fixed by academic rank and seniority. All professors of the same academic rank and seniority receive the same salary, regardless of the university that employs them. The information on individual salaries is unavailable but the salaries ranges for rank and seniority are published. Thus we approximate the salary for each individual as the average of their academic rank.

---

[6] For application of non-parametric methods to productivity measurements of Italian universities see Bonaccorsi and Daraio, 2003; Bonaccorsi et al. 2006; Abramo et al., 2008; Abramo et al., 2011b.

**3.2 Productivity indicators**

Based on the assumptions described above, the yearly average productivity over a certain period for researchers in a university in a particular SDS *S* is named Fractional Scientific Strength, $FSS_s$, and calculated as follows:

$$FSS_S = \frac{1}{S_{RS}} \sum_{i=1}^{N} \frac{c_i}{\bar{c}_i} f_i$$

---

[7] Although the overall coverage of the two databases does differ significantly, evidence suggests that, with respect to comparisons at large scale level in the hard sciences, the use of either source yields similar results (Archambault et al., 2009).

[1]

Where:
$S_{RS}$ = total salary of the research staff of the university in the SDS, in the observed period;
$N$ = number of publications of the research staff in the SDS of the university, in the period of observation;
$c_i$ = citations received by publication $i$;
$\bar{c}_i$ = average citations received by all cited publications of the same year and subject category of publication $i$;
$f_i$ = fractional contribution of researchers in the SDS of the university. Individual fractional contribution equals the inverse of the number of authors, in those fields where authors appear in alphabetical order in the byline, but assumes different weights in the life sciences, where we give different weights to each co-author according to their order in the byline and the character of the co-authorship (intra-mural or extra-mural)[8].

Based on $FSS_s$, we elaborate university ranking lists in each SDS on a percentile scale of 0-100 (worst to best) for comparison of absolute values of indicators relevant for all Italian universities active in the same SDS.

At UDA level we aggregate productivity measures of SDSs in the UDA, standardizing them to national average and weighting for the relative salary cost of the SDS. In this way, we take account of the varying intensity of productivity for the SDSs, avoiding the typical distortion of measures at the aggregate level.

The yearly average productivity $FSS_U$ of a university in a specific UDA $U$, is:

$$FSS_U = \sum_{k=1}^{M} \frac{FSS_{S_k}}{\overline{FSS_{S_k}}} \frac{S_{RS_k}}{S_{RS_U}}$$

[2]

With:
$M$ = number of SDSs of the university in the UDA $U$;
$\overline{FSS_{S_k}}$ = weighted[9] average $FSS_S$ of all universities with productivity above 0 in the SDS $k$.
$S_{RS_k}$ = total salary of the research staff of the university in the SDS $k$, in the observed period;
$S_{RS_U}$ = total salary of the research staff of the university in the UDA $U$, in the observed period.

Based on $FSS_U$, we elaborate universities ranking lists in each UDA on a percentile scale of 0-100 (worst to best) for comparison of absolute values of indicators relevant for all Italian universities active in the same UDA.

The same approach is used to measure overall yearly average productivity at university level, $FSS$:

---

[8]If first and last authors belong to the same university, 40% of the contribution is assigned to each of them; the remaining 20% is divided among all other authors. If the first two and last two authors belong to different universities, 30% of the contribution is assigned to first and last authors; 15% is attributed to second and last author but one; the remaining 10% is divided among all others. The weighting values were assigned following advice from senior Italian professors in the life sciences. The values could be changed to suit different practices in other national contexts.
[9]Weighting represents the relative size (in terms of cost of labor) of the SDS of each university.

$$FSS = \sum_{k=1}^{Z} \frac{FSS_{S_k}}{\overline{FSS_{S_k}}} \frac{S_{RS_k}}{S_{RS}}$$

[3]

With:
$Z$ = number of SDSs of the university;
$S_{RS}$ = total salary of the research staff of the university, in the observed period;
All other variables are as above.

**3.3 Dataset and sources**

Data on research staff of each university and their SDS classification are extracted from the database on Italian university personnel, maintained by the MIUR[10]. The bibliometric dataset used to measure productivity is extracted from the Italian Observatory of Public Research (ORP)[11], a database developed and maintained by the authors and derived under license from the Thomson Reuters' WoS. Beginning from the raw data of the WoS, and applying a complex algorithm for reconciliation of the author's affiliation and disambiguation of the true identity of the authors, each publication is attributed to the university scientist or scientists that produced it (D'Angelo et al., 2011).

Our dataset includes all universities active in the hard sciences. To ensure the representativity of publications as proxy of the research output, we include in the analysis only those SDSs where at least 50% of researchers produced at least one WoS-indexed publication in the period 2004-2008. The dataset is summarized in Table 1 in terms of SDSs, university, research staff, publications, and citations for each UDA.

*Table 1: Research staff and scientific portfolio of Italian universities in 2004-2008, by UDA*

| UDA | SDSs | Universities | Research staff | Publications* | Citations |
|---|---|---|---|---|---|
| Mathematics and computer sciences | 9 | 63 | 3,515 | 13,955 | 33,697 |
| Physics | 8 | 61 | 2,873 | 23,375 | 155,919 |
| Chemistry | 11 | 59 | 3,603 | 24,570 | 168,723 |
| Earth sciences | 12 | 48 | 1,439 | 4,641 | 20,422 |
| Biology | 19 | 67 | 5,785 | 27,985 | 221,029 |
| Medicine | 47 | 56 | 12,196 | 50,338 | 407,189 |
| Agricultural and veterinary sciences | 28 | 48 | 3,153 | 10,168 | 39,672 |
| Industrial and information engineering | 42 | 68 | 5,489 | 32,188 | 79,421 |
| Total | 176 | 77 | 38,053 | 164,632** | 975,402** |

\* *Number of publications authored by at least one academic scientist of the UDA.*
\*\* *The value differs from the column total due to multiple counts for publications by co-authors belonging to different UDAs.*

**4. Results and analysis**

We propose the following analysis to verify if universities that place in the top national ranks for overall productivity also succeed at placing in the top for the individual disciplines of their activity. We will prepare ranking lists at two levels:

---
[10]http://cercauniversita.cineca.it/php5/docenti/cerca.php, last accessed on July 1, 2013
[11]www.orp.researchvalue.it, last accessed on July 1, 2013

- One for overall productivity at university level as defined in [3], and for reasons of significance including only those universities (61 in all) with a research staff of not less than 20 individuals in the SDSs that we are investigating.
- One for each UDA, for productivity of universities at UDA level as defined in [2], and including only those universities with a research staff of not less than 10 in each UDA.

The comparison of these ranking lists (Section 4.1) will then permit us to respond to our research question.

Similarly, descending in level, we can also verify if the universities at the top in an individual UDA then also top the rankings for the individual constituent SDSs. For reasons of space, in reporting on this analysis (Section 4.2) we will present only the example of the Physics UDA.

For each ranking list we classify the university placing in terms of merit, corresponding to quintiles of productivity: A, B, C, D, E, with A the best and E the worst.

**4.1 Overall productivity vs UDAs productivity**

The 61 universities analyzed vary in the scope of their disciplines (Table 2). Only 14 universities are active in all eight of the UDAs considered; another 11 are active in seven disciplines. Seventeen universities are active in less than four UDAs, including four that are active in two UDAs and four in a single UDA. The area with the least number of active universities is Agriculture and veterinary sciences, with only 27. At the opposite extreme, Biology and Mathematics both have 49 active universities. In specific regard to the research question, a first observation concerns the lack of connection between the number of disciplines present in a university and its overall productivity. We find three generalist universities in the group of 12 for class A and another three in the group of 13 for class E; similarly, examining the universities with less than 5 active UDAs, we find 8 in class A but also 4 in class E.

*Table 2: Bibliometric productivity classes (A the best, E worst) for Italian universities (data 2004-2008)*

| ID | Mathematics and computer sciences | Physics | Chemistry | Earth Sciences | Biology | Medicine | Agriculture and Veterinary sciences | Industrial and information engineering | N. of UDAs | Class. UNIV | Stand. mutual variability index-R | ID | Mathematics and computer sciences | Physics | Chemistry | Earth Sciences | Biology | Medicine | Agriculture and Veterinary sciences | Industrial and information engineering | N. of UDAs | Class. UNIV | Stand. mutual variability index-R |
|---|---|---|---|---|---|---|---|---|---|---|---|---|---|---|---|---|---|---|---|---|---|---|---|
| UNIV_1 | - | - | - | - | A | A | - | - | 2 | A | 0 | UNIV_32 | C | A | B | E | D | C | - | B | 7 | C | 0.607 |
| UNIV_2 | A | A | - | - | - | - | - | - | 2 | A | 0 | UNIV_33 | D | A | D | B | B | C | C | C | 8 | C | 0.455 |
| UNIV_3 | - | - | - | - | - | - | - | A | 1 | A | N.A. | UNIV_34 | A | D | C | D | A | D | D | A | 8 | C | 0.616 |
| UNIV_4 | - | - | - | - | - | A | - | - | 1 | A | N.A. | UNIV_35 | B | D | A | - | B | C | - | C | 6 | C | 0.525 |
| UNIV_5 | A | A | - | - | - | - | - | A | 3 | A | 0 | UNIV_36 | E | B | E | - | A | B | D | E | 7 | C | 0.750 |
| UNIV_6 | A | B | - | - | A | - | - | - | 3 | A | 0.250 | UNIV_37 | C | D | A | C | C | D | - | D | 7 | D | 0.429 |
| UNIV_7 | - | - | D | - | A | - | A | - | 3 | A | 0.750 | UNIV_38 | E | E | - | - | D | C | C | - | 5 | D | 0.450 |
| UNIV_8 | A | - | - | - | A | B | A | - | 4 | A | 0.188 | UNIV_39 | A | E | C | C | C | D | - | D | 7 | D | 0.536 |
| UNIV_9 | C | C | A | A | A | A | - | A | 7 | A | 0.357 | UNIV_40 | C | - | D | - | C | - | - | - | 3 | D | 0.250 |
| UNIV_10 | B | D | A | B | B | A | A | A | 8 | A | 0.402 | UNIV_41 | - | - | - | - | E | D | B | - | 3 | D | 0.750 |
| UNIV_11 | A | A | - | - | C | A | - | C | 5 | A | 0.450 | UNIV_42 | D | E | C | C | C | D | A | B | 8 | D | 0.549 |
| UNIV_12 | D | D | C | A | B | A | A | E | 8 | A | 0.710 | UNIV_43 | C | E | B | B | E | C | E | B | 8 | D | 0.603 |
| UNIV_13 | A | B | E | - | B | - | - | A | 5 | B | 0.675 | UNIV_44 | E | C | C | C | D | D | B | D | 8 | D | 0.402 |
| UNIV_14 | B | B | A | D | B | - | A | - | 7 | B | 0.429 | UNIV_45 | - | - | E | B | D | - | - | - | 3 | D | 0.750 |
| UNIV_15 | C | C | B | B | B | A | B | B | 8 | B | 0.254 | UNIV_46 | E | C | E | D | C | C | - | E | 7 | D | 0.429 |
| UNIV_16 | B | E | D | - | B | B | B | B | 7 | B | 0.464 | UNIV_47 | - | - | - | - | D | E | C | - | 3 | D | 0.500 |
| UNIV_17 | - | - | - | - | - | - | B | - | 1 | B | N.A. | UNIV_48 | - | - | - | - | E | D | - | - | 2 | D | 0.375 |
| UNIV_18 | C | A | C | - | - | - | - | C | 4 | B | 0.375 | UNIV_49 | D | B | C | B | C | E | C | D | 8 | E | 0.455 |
| UNIV_19 | B | B | E | A | A | - | - | D | 6 | B | 0.725 | UNIV_50 | E | - | - | - | - | - | - | E | 2 | E | 0 |
| UNIV_20 | A | A | A | E | - | - | - | C | 5 | B | 0.750 | UNIV_51 | C | C | D | D | - | - | E | E | 6 | E | 0.400 |
| UNIV_21 | D | C | B | C | A | B | A | - | 7 | B | 0.500 | UNIV_52 | E | A | D | D | E | - | E | - | 6 | E | 0.600 |
| UNIV_22 | B | A | B | - | B | C | - | - | 5 | B | 0.300 | UNIV_53 | D | B | E | D | E | E | E | C | 8 | E | 0.469 |
| UNIV_23 | A | D | B | A | B | A | D | D | 8 | B | 0.603 | UNIV_54 | C | E | A | - | E | E | - | E | 6 | E | 0.650 |
| UNIV_24 | B | D | A | A | B | B | D | C | 8 | B | 0.522 | UNIV_55 | - | - | D | - | E | E | D | - | 4 | E | 0.250 |
| UNIV_25 | - | - | - | - | - | - | - | B | 1 | C | N.A. | UNIV_56 | E | C | E | - | D | D | - | E | 6 | E | 0.350 |
| UNIV_26 | B | B | B | - | C | - | - | C | 5 | C | 0.225 | UNIV_57 | D | - | - | - | - | - | E | E | 3 | E | 0.250 |
| UNIV_27 | - | - | - | E | E | - | - | A | 3 | C | 1 | UNIV_58 | E | C | A | D | D | E | C | B | 8 | E | 0.629 |
| UNIV_28 | D | - | E | A | D | B | - | - | 5 | C | 0.750 | UNIV_59 | E | E | D | E | D | E | - | D | 7 | E | 0.196 |
| UNIV_29 | B | E | B | C | C | B | - | A | 7 | C | 0.536 | UNIV_60 | D | C | E | E | E | E | E | D | 8 | E | 0.295 |
| UNIV_30 | B | D | C | - | A | C | - | - | 5 | C | 0.525 | UNIV_61 | D | - | - | E | E | - | - | E | 4 | E | 0.188 |
| UNIV_31 | E | E | - | E | - | - | - | B | 4 | C | 0.563 | | | | | | | | | | | | |

*4.1.1 Analysis of top and bottom universities*

An analysis of the top and bottom classes for overall productivity reveals that among the 12 universities in class A, the first five are at the top in all the UDAs where they are active. Four of these operate only in two UDAs that are closely related (e.g. Physics and Mathematics or Biology and Medicine). However this group of five includes two smaller Scuole Superiori and two private universities specializing in biomedicine. Beginning with UNIV_6, but still among the top universities for overall productivity, there is a continuous presence of UDAs that do not achieve level A. In UNIV_7, the Chemistry UDA places in Class D, the second last national quintile. For the three generalist universities (UNIV_9, UNIV_10, UNIV_12), performance is anything but uniform: in UNIV_9 five "excellent" UDAs are paired with two that fall in class C (Mathematics and computer sciences, Physics). For UNIV_10, four out of eight UDAs are non-excellent, including one in class D (Physics). UNIV_12, the last of the top group, shows a number of mediocre UDAs, with Physics and Mathematics and computer science UDAs in the second last quintile and Industrial and information engineering UDA actually placing in the bottom class.

Examining the rankings tail composed of the 13 universities in class E for overall productivity, we then find three excellent UDAs: Physics (UNIV_52) and Chemistry (UNIV_54, and UNIV_58). There are also a number of UDAs with performance placing in class B. One of the generalist institutions (UNIV_49) is particularly notable: this university is active in all eight UDAs, achieving good positions (class B) in Earth sciences and Physics and intermediate standing (class C) in Agriculture and veterinary sciences, Biology and Chemistry.

Overall, we do not detect any clear stratification of universities on the basis of their research performance. Although appearing as a scattered minority, some excellent UDAs (class A) occur even in the tail of the general university ranking. At the same time, the universities that achieve overall excellent productivity include UDAs that do not reach the general standard of performance for their institutions.

*4.1.2 Performance variability of UDAs within universities*

As the measure of the potential mixed levels of performance of the UDAs active in each university, we calculate the index of mutual variability of the classes (A to E), symbolized Δ. This represents the average of the differences, in absolute value, of all possible pairs without repetition. In formulae:

$$\Delta = \frac{\sum_{i \neq j=1}^{n} |x_i - x_j|}{n(n-1)}$$

[4]

Where:
$x_i$ = quintile of performance for the UDA $i$ (1 the best, 5 worst),
$n$ = number of UDAs active at the university.

We then examine the results for the standardized value R, obtained by dividing Δ by the maximum of the Δ values registered for the 57 universities[12]. By definition, R assumes nil value in correspondence with nil dispersion in the performance of the UDAs, while an R value of one corresponds to the situation of maximum dispersion in UDA performance.

Among the 57 universities there are four cases where R value is nil, however these concern highly specialized institutions with only two active UDAs (UNIV_1, UNIV_2, UNIV_50), or a maximum of three UDAs (UNIV_5). The maximum dispersion of performance (R=1) is registered for UNIV_27, which places in the bottom performance percentile (E) in Earth Sciences and Biology and the top (A) in Industrial and information engineering.

Overall, we observe a Gaussian type of distribution for the R values, with the mode at the 0.4-0.6 inter-quintile, accounting for 36.8% of total observations. The median for the R distribution is 0.455, which is the specific case for two generalist universities: UNIV_33, where UDA performance swings from A (Physics) to D (Chemistry, Mathematics and computer sciences), and UNIV_49, with two UDAs in class B (Physics, Earth sciences) and one in class E (Medicine).Taking a closer look at the other 23 generalist universities (considered as having 7 or 8 active UDAs),we note that there are 14 cases where the value of R is above the median. However variability in performance does not result as correlated to the number of active UDAs: in fact the coefficient of correlation between the values in columns 10 and 12 (Table 2) is 0.28.

---
[12] Four of the original 61 universities under observation are excluded because they are active in only one UDA.

*4.1.3 Analysis by class*

Still examining the heterogeneous performance by the UDAs in individual universities, Table 3 presents the statistics from a series of analyses by group. For the universities in each quintile for overall productivity, the table shows the average distribution of the performance by the active UDAs. For example, the second line in column 3 shows that for the 12 universities in class A by overall productivity, the average percentage of UDAs with productivity in class A is 77%, while the remaining 23% of their UDAs obviously register performance below "top". The data that most clearly demonstrate the lack of consistency in performance are the values along the principle diagonal (highlighted in yellow)[13]. For the 12 universities placing in overall productivity class B, there are an average of 27% of UDAs in top class. For this same overall group, 15% of UDAs register productivity in class C, 10% in class D and 6% in class E. Among the group of 12 universities that register overall productivity around the national median (class C), we find that 15% of UDAs are excellent (class A).

*Table 3: Distributions of UDA productivity (by class, A the best, E worst) for universities grouped by overall productivity*

| Overall productivity class | N. of universities | Average of A class UDAs | Average of B class UDAs | Average of C class UDAs | Average of D class UDAs | Average of E class UDAs |
|---|---|---|---|---|---|---|
| A | 12 | 77% | 9% | 7% | 6% | 1% |
| B | 12 | 27% | 42% | 15% | 10% | 6% |
| C | 12 | 15% | 32% | 18% | 15% | 19% |
| D | 12 | 3% | 11% | 31% | 29% | 26% |
| E | 13 | 4% | 4% | 12% | 27% | 54% |

**4.2 UDA productivity vs SDSs productivity**

In this section we will compare the rank standings for productivity by universities active in an individual UDA to their rankings in the SDSs that compose it, to understand if the universities at the top in a given discipline are also top in the individual subfields. For reasons of space we present only the example of the Physics UDA, with its eight constituent SDSs[14]. For each of the 43 universities with at least 10 faculty in the Physics

---

[13] The apparently greater homogeneity within the universities of the top and bottom classes is partially due to the fact that performance in the individual disciplines can be equal to or less (for those in top class) and equal to or greater (for bottom class), while it can be greater, equal or lesser for the other classes.

[14] FIS/01: Experimental Physics; FIS/02: Theoretical Physics, Mathematical Models and Methods; FIS/03: Material Physics; FIS/04: Nuclear and Subnuclear Physics; FIS/05: Astronomy and Astrophysics;

UDA, Table 4 shows the performance class for the overall UDA, for the SDSs active within that particular UDA, and the standardized mutual variability index R.

*Table 4: Productivity classes (A the best, E worst) for Italian universities in Physics: SDSs analysis (2004-2008 data)*

| ID | CLASS. UDA | FIS/01 | FIS/02 | FIS/03 | FIS/04 | FIS/05 | FIS/06 | FIS/07 | FIS/08 | TOT SDS | Research staff | Standardized mutual variability index-R |
|---|---|---|---|---|---|---|---|---|---|---|---|---|
| UNIV_2  | A | -  | C | A | A | A | -  | A | -  | 5 | 31  | 0.267 |
| UNIV_5  | A | A  | A | C | A | -  | A | -  | -  | 5 | 44  | 0.267 |
| UNIV_11 | A | A  | - | A | -  | -  | -  | B | -  | 3 | 14  | 0.222 |
| UNIV_18 | A | A  | - | A | D | -  | -  | -  | -  | 3 | 52  | 0.667 |
| UNIV_20 | A | A  | A | C | A | -  | -  | -  | -  | 4 | 41  | 0.333 |
| UNIV_22 | A | A  | B | A | C | A | -  | E | -  | 6 | 24  | 0.6   |
| UNIV_32 | A | C  | B | B | -  | D | A | A | -  | 6 | 46  | 0.467 |
| UNIV_33 | A | A  | E | B | E | -  | D | E | A | 7 | 41  | 0.730 |
| UNIV_52 | A | C  | E | A | E | -  | A | A | -  | 6 | 23  | 0.755 |
| UNIV_6  | B | E  | D | A | E | C | -  | -  | -  | 5 | 17  | 0.667 |
| UNIV_13 | B | A  | C | D | B | E | B | C | -  | 7 | 69  | 0.540 |
| UNIV_14 | B | B  | A | C | C | B | D | B | D | 8 | 66  | 0.417 |
| UNIV_19 | B | C  | A | A | E | B | D | A | -  | 7 | 51  | 0.635 |
| UNIV_26 | B | D  | B | B | -  | -  | -  | D | -  | 4 | 44  | 0.445 |
| UNIV_36 | B | A  | - | -  | -  | -  | -  | D | -  | 2 | 13  | 1     |
| UNIV_49 | B | B  | A | A | D | D | C | D | E | 8 | 183 | 0.595 |
| UNIV_53 | B | A  | D | D | E | C | E | A | C | 8 | 67  | 0.619 |
| UNIV_9  | C | B  | E | D | A | D | D | B | E | 8 | 43  | 0.583 |
| UNIV_15 | C | C  | C | C | B | B | -  | C | E | 7 | 135 | 0.349 |
| UNIV_21 | C | C  | B | E | C | B | C | B | -  | 7 | 89  | 0.381 |
| UNIV_44 | C | B  | C | D | B | A | C | A | B | 8 | 160 | 0.405 |
| UNIV_46 | C | B  | E | B | D | -  | D | A | A | 7 | 91  | 0.635 |
| UNIV_51 | C | C  | E | -  | -  | -  | A | -  | -  | 3 | 12  | 0.889 |
| UNIV_56 | C | C  | A | B | -  | -  | -  | C | -  | 4 | 12  | 0.389 |
| UNIV_58 | C | B  | E | D | D | -  | -  | E | A | 6 | 51  | 0.645 |
| UNIV_60 | C | D  | A | D | A | E | -  | D | B | 7 | 86  | 0.635 |
| UNIV_10 | D | E  | D | E | E | A | B | C | B | 8 | 123 | 0.631 |
| UNIV_12 | D | D  | D | B | B | D | A | C | E | 8 | 108 | 0.536 |
| UNIV_23 | D | D  | B | B | B | E | B | B | -  | 7 | 95  | 0.413 |
| UNIV_24 | D | B  | D | C | A | E | C | D | A | 8 | 112 | 0.583 |
| UNIV_30 | D | B  | D | E | -  | -  | E | E | -  | 5 | 17  | 0.467 |
| UNIV_34 | D | C  | E | E | -  | -  | -  | B | C | 5 | 67  | 0.533 |
| UNIV_35 | D | D  | A | D | B | D | B | E | -  | 7 | 89  | 0.571 |
| UNIV_37 | D | D  | C | D | D | A | -  | B | -  | 6 | 62  | 0.489 |
| UNIV_16 | E | E  | - | E | -  | -  | E | A | C | 5 | 18  | 0.667 |
| UNIV_29 | E | E  | - | E | -  | -  | -  | E | -  | 3 | 14  | 0     |
| UNIV_31 | E | E  | - | E | -  | -  | -  | -  | -  | 2 | 14  | 0     |
| UNIV_38 | E | E  | B | E | -  | -  | -  | D | -  | 4 | 12  | 0.555 |
| UNIV_39 | E | E  | D | C | C | C | E | C | E | 8 | 73  | 0.369 |
| UNIV_42 | E | E  | B | E | E | -  | E | D | C | 7 | 64  | 0.445 |
| UNIV_43 | E | E  | C | C | C | B | -  | C | -  | 6 | 43  | 0.333 |
| UNIV_54 | E | D  | E | C | D | E | B | E | E | 8 | 51  | 0.417 |
| UNIV_59 | E | D  | C | B | C | C | -  | E | -  | 6 | 49  | 0.4   |

Ten of the universities are active in all eight of the Physics SDSs, and further ten are

---

FIS/06: Physics for Earth and Atmospheric Sciences; FIS/07: Applied Physics (Cultural Heritage, Environment, Biology and Medicine); FIS/08: Didactics and History of Physics.

active in seven. At the other end of the range, six universities are active in less than four UDAs, with two being active in only a pair of SDSs. Analyzing the distribution of SDSs within the individual universities we observe that FIS/01 (Experimental physics) is the most frequently present (42 universities out of 43), closely followed by FIS/03 Materials physics (41 out of 43), while FIS/08 (History and teaching of physics) is the least represented (18 universities).

In the nine universities at the top (class A) for overall productivity, we observe a full six SDSs in class E: FIS/07 in UNIV_22; FIS/02, FIS/04 and FIS/07 in UNIV_33; FIS/02 and FIS/04 in UNIV_52. At the tail of the UDA ranking, among the nine universities in the last quintile (class E) there is one excellent SDS (FIS/07 in UNIV_16) and a further five cases of SDSs in class B (FIS/02 in UNIV_38 and UNIV_42; FIS/03 in UNIV_59; FIS/05 in UNIV_43; FIS/06 in UNIV_54).

Overall, the situation remains analogous to that seen in the previous section: descending a level in the analysis of productivity, we now observe that the universities at the top in a discipline (in this example, the Physics UDA) are often not at the top for the single fields that compose the UDA. Excellent SDSs occur in universities with UDA productivity that is below top, and vice versa. In general, it emerges that there is strong heterogeneity in the performance of the fields within the same discipline at an individual university. The standardized mutual variability index (R) data provide confirmation, showing a clearly Gaussian distribution, with the modal value at the inter-quintile 0.4-0.6, accounting for 42% (18 out of 43) total observations. The maximum value is registered for UNIV_36, which is active in only two SDS (FIS/01, FIS/07), but with performance at almost opposite ends of the range (in this case A and D). UNIV_51 also shows notable variability in performance, with one SDS in class A (FIS/06), one in class E (FIS/02) and an intermediate one in class C (FIS/01). Among the 20 generalist universities (7 or 8 active SDSs), 12 show a value for R that is greater than 0.5 and six show a value over 0.6. The university with the largest research staff (UNIV_49) is active in all eight SDSs and shows very high variability in performance (R=0.595) spanning the entire spectrum of five classes. The same occurs for UNIV_10, another large generalist university, with an R value of 0.631. We also note that the dispersion of performance among SDSs in a UDA does not seem related either to the size of the research staff (coefficient of correlation between columns 12 and 13, equaling 0.03), or to the actual number of SDSs active in the UDA (coefficient of correlation between columns 11 and 13, equal to 0.13).

The heterogeneity in performance by the fields is also shown by the data in Table 5, which gives the statistics for an analysis by group. For the universities of each class (A to E) for productivity in Physics, the table indicates the average distribution of performance for the active SDSs in each institution. For example, the second line shows that for the nine "class A" universities for UDA productivity, the percentage of SDSs with productivity "not class A" is 44%, with this 44% distributed almost equally among the other classes. The most evident demonstration of the non-homogeneity of performance among the SDSs is the series of values along the principal diagonal (highlighted in yellow). Again, among the nine universities in class C for productivity in Physics (the median position in national ranking for this UDA), 18% of their SDSs have top productivity, while another 16% have productivity in the bottom class. We can conclude that, as seen for the level of UDAs, performance distribution at field level is substantially heterogeneous. Similar results were obtained from the analyses conducted for other UDAs.

*Table 5: Distribution of SDSs by productivity classes (A the best, E worst) for universities in Physics*

| UDA productivity class | N. of universities | Average of A class SDSs | Average of B class SDSs | Average of C class SDSs | Average of D class SDSs | Average of E class SDSs |
|---|---|---|---|---|---|---|
| A | 9 | 56% | 11% | 13% | 7% | 13% |
| B | 8 | 22% | 18% | 18% | 27% | 14% |
| C | 9 | 18% | 26% | 21% | 19% | 16% |
| D | 8 | 11% | 28% | 13% | 26% | 22% |
| E | 9 | 2% | 10% | 29% | 12% | 47% |

## 5. Conclusions

Many countries clearly look to the model of elite Anglo-Saxon universities as inspiration for national policies designed to strengthen scientific infrastructure and higher education systems. In the pursuit of this objective, it appears essential the selective public funding of the higher education system, a choice that implies a strong break with the past in the case of non-competitive systems. In such systems, the historic dispersion of public resources has actually undermined any opportunity for differentiation among universities and the development of top universities that can compete at the international level. The current study demonstrates that this dispersion also occurs within the individual universities. "Excellent" disciplines occur even in universities in the tail of national rankings for overall productivity, while single universities with overall excellent productivity repeatedly include disciplines that are below their institution's overall performance.

In addition to these observations we have added a further level of analysis at a lower level, examining the distribution of performance for the fields in the same discipline: in the individual universities, the excellent fields coexist with mediocre fields.

This significant internal variability of performance in single institutions is likely not only a phenomenon of the Italian academic system, but more broadly of the majority of non-competitive higher education systems, determining the absence of any universities with a critical mass of excellence sufficient to aspire to the ranks of "elite" universities.

We pose the question of whether the scattered capacities to produce excellent results could be better employed by concentrating such capacities, in terms of the available human and material resources, in a much smaller number of universities. With very low investment, it would seem possible to quickly create the type of elite universities that other competitive systems have produced over decades of time: universities capable of attracting the best faculty, students and public and private capital, and of providing much greater economic benefit than the present universities, with their high dispersion of internal performance.

The authors sincerely hope that decision makers in undifferentiated non-competitive higher education systems will consider this policy option, particularly in the current times of economic crisis, which call for courageous and decisive choices for the future good of our nations.